\documentclass[prl, notitlepage, superscriptaddress,  reprint]{revtex4-1}
\usepackage{enumerate}
\usepackage{amssymb}
\usepackage{amsmath}
\usepackage{braket}
\usepackage{graphicx}
\usepackage[usenames,dvipsnames]{color}
\usepackage[colorlinks,bookmarks=false,citecolor=NavyBlue,linkcolor=OliveGreen,url
color=blue]{hyperref}
\usepackage{tensor}
\usepackage{tikz}
\definecolor{myorange}{RGB}{199.24, 87.48, 47.80}
\usetikzlibrary{decorations.markings}
\newcommand{\be}{\begin{equation}}
\newcommand{\ee}{\end{equation}}
\newcommand{\ba}{\begin{aligned}}
\newcommand{\ea}{\end{aligned}}
\newcommand{\1}{{\rm I}}
\newcommand{\bw}{\begin{widetext}}
\newcommand{\ew}{\end{widetext}}

\def\doi{http://dx.doi.org/}

\begin{document}
\title{Determination of the Nonequilibrium Steady State Emerging from a Defect}
\author{Bruno Bertini}
\affiliation{SISSA and INFN, via Bonomea 265, 34136, Trieste, Italy}
\author{Maurizio Fagotti}
\affiliation{D\'epartement de Physique, \'Ecole Normale Sup\'erieure / PSL Research
University, CNRS, 24 rue Lhomond, 75005 Paris, France}
\begin{abstract}
We consider the non-equilibrium time evolution of a translationally invariant state under a Hamiltonian with a localized defect. We discern the situations where a light-cone spreads out from the defect and separates the system into regions with macroscopically different properties. We identify the light-cone and propose a procedure to obtain a (quasi-)stationary state describing the late time dynamics of local observables. As an explicit example, we study the time evolution generated by the Hamiltonian of the transverse-field Ising chain with a local defect that cuts the interaction between two sites (a quench of the boundary conditions alongside a global quench). We solve the dynamics exactly and show that the late time properties can be obtained with the general method  proposed.
\end{abstract}
\maketitle

In an out-of-equilibrium many-body quantum system, some observables can relax even if the system is isolated. The last decade has brought fresh insights into this counter-intuitive phenomenon, already addressed almost one hundred years ago by J. von Neumann~\cite{vN}.  This renewed interests started with a series of groundbreaking experiments~\cite{exp,kww-06}. Arguably, the most inspiring of these has been the \emph{Quantum Newton's Cradle}~\cite{kww-06}, which motivated copious theoretical research on the role of dimensionality and conservation laws out-of-equilibrium \cite{EF:review, P:review, quench,GE15}. The theoretical picture emerged is that in translationally invariant systems the expectation value of any local observable approaches a stationary value. This can be computed in the effective stationary state with maximum entropy under the constraints of the relevant integrals of motion. 
Since integrable models have infinitely many local conservation laws, which do affect the dynamics of local observables, the stationary state emerging in those systems is very different from standard statistical ensembles. It was called generalized Gibbs ensemble (GGE)~\cite{Rigol07} and is characterized by the expectation values of local and quasilocal conservation laws~\cite{Doyon,P:review, EF:review}. 

In practice, the set of the local charges is very sensitive to global perturbations and real systems are never exactly integrable. Thus, in actual experiments relaxation to a GGE is only a property of (possibly very long) intermediate times, which precede the onset of  thermalization~\cite{thermalization}. The study of integrability breaking perturbations 
 led to new theoretical concepts, such as prethermalization~\cite{prethermalization, prerelaxation} and pre-relaxation~\cite{prerelaxation}, which have been observed in experiments~\cite{LGS:review,prethermalizationexp}. These regimes occur at timescales such that the fastest degrees of freedom have already relaxed, but the approximate integrability of the model still plays a key role.

When translational invariance is broken, other forms of stationary behavior appear. 
The typical example is the evolution of inhomogeneous states obtained  joining together two chains at different temperatures~\cite{thermoelectricengine, BD:review, VM:review, twotemperatures, conjTT} or with different magnetizations \cite{twomagnetizations}, or any other different global property~\cite{otherdifferentglobalproperties}. Around the junction of the two chains a non-equilibrium steady state (NESS)~\cite{Ruelle,BD:nessCFT} emerges.  

Importantly, a NESS can be produced simply by a Hamiltonian defect $\hat d$ localized around a given position $x_0$, even if the initial state is homogeneous~\cite{Fdefect}. The mechanism for this is based on the fact that the set of the local conservation laws is also very sensitive to localized perturbations. Some charges $Q_j$ are simply deformed by the defect, \emph{i.e.} there are bounded operators $\delta Q_j$ localized around $x_0$ such that
\be
\ba
{}[H_0,Q_j]=0\, ,&&&&[H_1,Q_j+\delta Q_j]=0\, ,
\ea
\ee
where $H_0$ is the Hamiltonian of the ``clean'' model  and $H_1=H_0+\hat d$ is the Hamiltonian with the defect. In general, however, there are also charges $\tilde Q_j$ of $H_0$ for which such bounded operators do not exist, \emph{i.e.} they cannot be deformed into conserved operators for $H_1$. We qualify them as ``extinct''.
In contrast to the charges which are deformed, the extinct charges have the peculiarity that the expectation value of their commutator with the defect can remain nonzero for arbitrarily large time
\be\label{eq:AQ}
\mathcal A_{\tilde Q}=\lim_{t\rightarrow\infty }\mathrm{tr}\bigl[\rho_t\, i[\tilde Q,\hat d]\bigr]\neq 0\, ,
\ee 
where $\rho_t$ is the time evolving state. This is sufficient to ensure that the limit of infinite time \emph{does not} commute with the limit of infinite distance from the defect~\cite{Fdefect}:  a light-cone spreads out from $x_0$ separating the system into regions with \emph{macroscopically different properties}.  

We note that some charges of $H_1$ can also result from the merging of two distinct charges of the clean model, in the sense that their densities in the bulk are different on the two sides of the defect. 
For the sake of simplicity we will consider defects that cut the chain in two and focus on one side; this allows us to classify the charges simply as deformed or extinct.

In this Letter we first show that, in generic quantum spin chains, far enough from $x_0$ (outside the light-cone) the dynamics are effectively governed by the clean Hamiltonian $H_0$. Then, we propose a procedure to construct the effective stationary states that, at sufficiently large time within the light-cone, become locally equivalent to $\rho_t$.
To illustrate our ideas, we consider the prototypical quantum quench where the condition \eqref{eq:AQ} can be fulfilled: a translationally invariant state on a chain that evolves under a Hamiltonian with free boundaries (similar protocols were considered in \cite{SM:toc,CVZ:14}). Indeed, in integrable models a free boundary induces the extinction of infinitely many local charges~\cite{openXYZ}.
We work out the time evolution and construct the NESS emerging from the boundaries for the particular case of the transverse-field Ising chain (TFIC).  
We identify a locally-quasi-stationary state (LQSS)  
which describes subsystems in the limit of large time and distance.
We show that  the NESS and even the full LQSS can be obtained using our general procedure.

\paragraph{The light-cone.}
We consider a generic spin-chain in an homogeneous state $\rho_0$ evolving with the Hamiltonian $H_1=H_0+\hat d$, where $H_0$ is translationally invariant and $\hat d$ acts nontrivially only in a finite region $D$. We indicate with $\mathcal O_A$ a local operator acting nontrivially only on $A$.  
Let $S$ be a subsystem that contains $A$ and $\bar S$ its complement. 
We define the projector of the operator $\mathcal O_A (t)=e^{i H_1 t}\mathcal O_A e^{-i H_1 t}$ to $S$ by
$
\mathcal O^{(S)}_A (t)=\mathrm{tr}_{\bar S}[\mathcal O_A(t)]\otimes \1_{\bar S}/\mathrm{tr}[\1_{\bar S}]
$
and  $\delta \mathcal O^{(S)}_A (t)=\mathcal O_A(t)-\mathcal O^{(S)}_A (t)$.
If $D\cap S=\emptyset$, the projected operator commutes with the defect.
We then find~\cite{f:1}
\be
\label{eq:ineq}
||\mathcal O_A(t)-e^{i H_0 t}\mathcal O_Ae^{-i H_0 t}||\leq  \int_{0}^t\mathrm d \tau||[\delta \mathcal O^{(S)}_A (\tau),\hat d]||\, ,
\ee
where $||\cdot||$ is the operator norm.
Using the Lieb-Robinson bound~\cite{LR72}, Ref.~\cite{bravyi06} obtained
\be\label{eq:boundLB}
||\delta \mathcal O^{(S)}_A (t)||\leq c e^{-\frac{\ell-v t}{\xi}}\, 
\ee
where $\ell$ is the (smallest) distance between $A$ and $\bar S$, $v$ is the Lieb-Robinson velocity and  $\xi$ and $c$ are constants. The largest value of $\ell$ compatible with $D\cap S=\emptyset$ is the distance $r$ between $D$ and $A$. For that distance, \eqref{eq:ineq} reads
\be\label{eq:boundLB1}
|| \mathcal O_A(t)- e^{i H_0 t}\mathcal O_Ae^{-i H_0 t}||\leq  2cv^{-1}\xi|| \hat d|| e^{-\frac{r}{\xi}}(e^{\frac{v t}{\xi}}-1).
\ee
In the \emph{space-time scaling limit} $r, t\rightarrow\infty$ at 
$\kappa =r/(vt)$ fixed, we thus find
\be\label{eq:Op}
\parallel \mathcal O_A(t)-e^{i H_0 t}\mathcal O_A e^{-i H_0 t}\parallel\rightarrow
0\qquad \kappa>1
\ee
\emph{i.e.}
the evolution under $H_1$ is equivalent to the evolution under $H_0$.
This is a proof of the very natural fact that a defect is irrelevant outside the  light-cone at $\kappa=1$.

In addition, for $\kappa>1$ the problem is reduced to a quantum quench in a system with translational invariance.   
Since \eqref{eq:Op} is in the limit of infinite time and $\mathcal O_A$ is local,  for $\kappa>1$ the state can be replaced by a GGE for the clean model described by $H_0$ \cite{EF:review}
\be\label{eq:GGE}
\rho_t \rightarrow \rho^{\textsc{gge}}=Z^{-1}e^{-\sum_i(\lambda_i Q_i+\tilde \lambda_i \tilde Q_i)}\, .
\ee  
Here the Lagrange multipliers $\lambda_i,\tilde\lambda_j$ can be fixed computing the integrals of motion in the initial state.

\paragraph{The NESS.}
The characterization of the NESS emerging from the defect is  more complicated. 
The first problem to face is its parametrization.  
It is reasonable to expect that, close to the defect, at late times the state becomes locally equivalent to a stationary state for $H_1$.  For a defect that cuts the chain in two, we focus on the right part and propose the ansatz of a GGE constructed \emph{only} with the deformed charges
\be\label{eq:NESS}
\rho^{\textsc{ness}}=Z^{-1}e^{-\sum\nolimits_i\mu_i (Q_i+\delta Q_i)}\, .
\ee
Since $\delta Q_i$  are (quasi)localized around $D$,  they do not affect observables acting far away from the defect. That is to say, 
in the limit $\lim_{r\rightarrow\infty}\lim_{t\rightarrow\infty}$ (order matters!) observables can be described by a simplified translationally invariant state:
\be\label{eq:barrho}
\rho^{\textsc{ness}}\rightarrow \rho^{\textsc{ness}}_+=Z^{-1}e^{-\sum\nolimits_i\mu_i Q_i}\, .
\ee
This is a stationary state for the clean model and \emph{is generically different from \eqref{eq:GGE}}.

Having specified the form of the NESS, we come to the second step: \emph{fixing the Lagrange multipliers}. 

One could be tempted to fix $\mu_i$ in \eqref{eq:NESS}
by imposing the  integrals of motion of the deformed charges. As we will show for the TFIC, this  gives an incorrect result and the explanation is simple:  in the thermodynamic limit an integral of motion per unit length is \emph{solely} determined by its density \emph{outside} the light-cone.

\paragraph{The invariants.}
In its simplest formulation, our strategy to fix $\mu_i$ is to find observables $I_{r}$ which can be approximated by local operators lying at distance $r$ from the defect and satisfy 
 \be\label{eq:invq}
\mathcal I=\!\!\lim_{t\rightarrow\infty, r=\kappa v t}\mathrm{tr}[\rho_t  I_r]=\mathrm{tr}[  \rho^{\textsc{gge}}  I_0]\qquad \forall\kappa>0\, .
\ee 
We then call $\mathcal I$ \emph{invariant}. 
We remind the reader that, for $\kappa>1$, \eqref{eq:invq} is true for any local observable (\emph{cf}. \eqref{eq:GGE}), so it is a nontrivial requirement only for $0<\kappa\leq 1$.
For defects preserving integrability, as those we consider,  it is reasonable to expect that the NESS describes a region growing proportionally to the time
\be \label{eq:hyp}
\lim_{\kappa\rightarrow 0^+}\lim_{t\rightarrow\infty,  r= \kappa v t}\mathrm{tr}[\rho_t I_r]= \lim_{r\rightarrow\infty}\lim_{t\rightarrow\infty}\mathrm{tr}[\rho_t I_r]\, .
\ee
On the right-hand side $\rho_t$ can be replaced by $\rho^{\textsc{ness}}_+$.
Taking $\kappa\rightarrow 0^+$ in \eqref{eq:invq} then gives
\be \label{eq:condition}
\mathrm{tr}[\rho_+^{\textsc{ness}}  I_r]=\mathrm{tr}[  \rho^{\textsc{gge}}   I_r]\, .
\ee 
Sufficiently many invariants allow to fix $\rho_+^{\textsc{ness}}$ and, by going back to \eqref{eq:barrho} from \eqref{eq:NESS}, also the NESS~\cite{f:2}. 

In the presence of interactions, one can not generally define invariants based on observables $I_r$ independent of the ratio $r/vt$. 
As we will show in the example of the TFIC, the invariants can be determined with the help of a semiclassical picture of quasi-particles produced after the quench~\cite{cc-05}. With some modifications, this picture holds true also in interacting integrable models.

\paragraph{Example: the transverse-field Ising chain.}
We now concentrate on the evolution under the Hamiltonian
\be
H_{\mathtt s}^{(h)}=-\frac{J}{2}\sum\nolimits_{j}\Bigl(\sigma_j^x\sigma_{j+1}^x+h\sigma_j^z\Bigr)+\frac{J\mathtt s}{2}\sigma_0^x\sigma_{1}^x\, ,
\label{Eq:IsingHam}
\ee
where $\sigma_j^\alpha$ are Pauli matrices acting nontrivially only on the $j$-th site and $j\in]-\frac{L}{2},\frac{L}{2}]$ ($L$ is even). We set $h>1$ and impose periodic boundary conditions $\sigma_{-L/2}^x\equiv \sigma_{L/2}^x$. For $\mathtt s=1$ the interaction between two sites is cut and two free boundaries appear. 
The Hamiltonian $H_1(h)$ can be mapped to a quadratic form of Majorana fermions $a_j^{x,y}$ satisfying $\{a_\ell^\alpha, a_n^\beta\}=2\delta_{\ell n}\delta_{\alpha\beta}$,  through the Jordan-Wigner (J-W) transformation  $a_{\ell}^{x,y}=(\prod_{j}^{\ell-1}\sigma_j^z)\sigma_\ell^{x,y}$. 
As explicit examples we will consider quenches from the  ground state of the Hamiltonian
\be\label{eq:Hini}
H_{\rm ini}^{(h_0,\Delta)}=H_0^{(h_0)}+\frac{J \Delta}{4}\sum\nolimits_{j}\sigma_j^x\sigma_{j+1}^y-\sigma_j^y\sigma_{j+1}^x\, .
\ee 
However, we will present results valid for any translationally invariant initial state that is Gaussian in the J-W fermions.

\paragraph{Exact solution of the dynamics.}

\begin{figure}[t]
\includegraphics[width=0.48\textwidth]{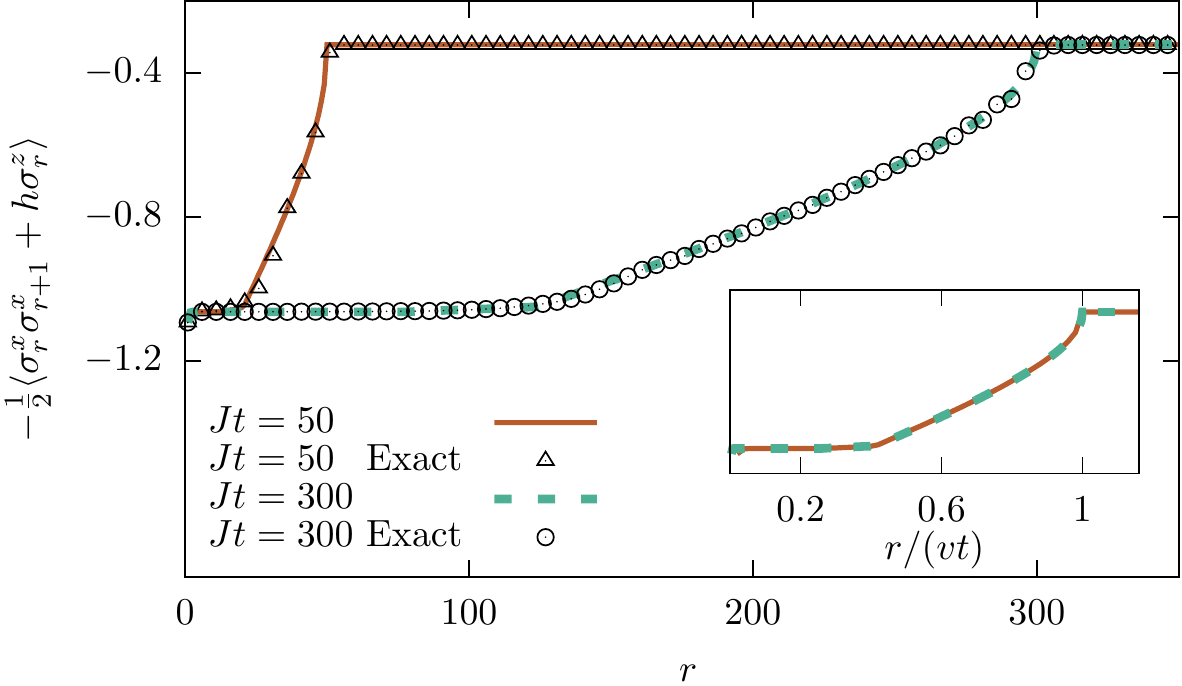}
\caption{The energy density as a function of the distance at two large times after the quench $H_{\rm ini}^{(1.5,4)}\rightarrow H_1^{(2)}$. Curves are the analytic predictions; symbols are exact numerical data for a finite chain with $L=1500$ sites. The inset shows that the density approaches a function of $\kappa=r/(vt)$.}\label{f:energy}
\end{figure}

Since both the pre-quench and the post-quench Hamiltonians are quadratic, by Wick's theorem all correlation functions can be reconstructed from the fermion two-point functions, which can be conveniently organized as a matrix
\be
\!\!\Gamma_{\ell n}(t)=\delta_{\ell n}\1-\begin{bmatrix}
 \mathrm{tr}[\rho_t a_{\ell}^x a_{n}^x ] & \mathrm{tr}[\rho_t a_{\ell}^x a_{n}^y ] \\
 \mathrm{tr}[\rho_t a_{\ell}^y a_{n}^x ] & \mathrm{tr}[\rho_t a_{\ell}^y a_{n}^y ] 
\end{bmatrix}.
\ee  
Here $\1$ is the 2-by-2 identity matrix. 
Using also that the initial state is translationally invariant, the correlation matrix at time $t$ can be recast in the  form 
\be
\Gamma_{\ell n}(t) = \frac{4}{L}\sum\nolimits_{q\in \mathcal S_p} \tilde{\mathbb{M}}_{\ell}^{ q}(t)
\Gamma_{ q}[{\tilde{\mathbb{M}}_{n}^{ q}}(t)]^\dag
\,.\label{Eq:Gammat}
\ee
Here $\Gamma_{q} $  is the Fourier transform of a block-row (also called symbol) of the correlation matrix of the initial state, which can be parametrized as $\Gamma_{q} =  f^{\textsc{i}}_q \1 + f^{\textsc{x}}_q \tilde\sigma^x_q+ f^{\textsc{y}}_q \tilde\sigma^y_q+ f^{\textsc{z}}_q \sigma^z
$.
The coefficients $ f^{\textsc{i}}_q, f^{\textsc{x}}_q, f^{\textsc{y}}_q,  f^{\textsc{z}}_q$ completely characterize the state and $f^{\textsc{i}}_q, f^{\textsc{x}}_q, f^{\textsc{z}}_q$  are odd functions of $q$, while $ f^{\textsc{y}}_q$ is even; $\tilde\sigma^\alpha_q=e^{i \theta_q \sigma^z/2} \sigma^\alpha e^{-i \theta_q \sigma^z/2}$, ${e^{i\theta_q}=J (h-e^{i q})/\varepsilon_q}$, and $\varepsilon_q=J \sqrt{1+h^2-2h\cos(q)}$. The matrix $\tilde{\mathbb{M}}_{n}^{\tilde q}(t)$ reads
\begin{gather}
\tilde{\mathbb{M}}_{n}^{ q}(t)=\sum_{j}\sum_{p\in \mathcal{S}_o}\mathbb{M}_{nj}(p,t) e^{- i  q j}
\label{Eq:Mtilde}\\
\mathbb{M}_{ij}(p,t)=
\begin{bmatrix}
 u_{i,p}u_{j,p}\cos(\varepsilon_p t) &u_{i,p}v_{j,p}\sin(\varepsilon_p t)  \\
 -v_{i,p}u_{j,p}\sin(\varepsilon_p t)  & v_{i,p}v_{j,p}\cos(\varepsilon_p t)
\end{bmatrix}\,,
\end{gather}
where $u_{n,p}= J \left(h \sin(n p)-\sin( (n-1) p)\right)/\mathcal{N}_p$, $v_{n,p}= \sin(n p) \varepsilon_p/\mathcal{N}_p$, and  $\mathcal{N}_p=(L\varepsilon_p^2+J^2h(h-\cos p))^{1/2}$.
The sums in \eqref{Eq:Gammat} and \eqref{Eq:Mtilde} run over distinct momenta in the sets $\mathcal S_p\in]-\pi,\pi[$ and $\mathcal S_o\in]0,\pi[$, quantized according to 
$q\in\mathcal{S}_p\Leftrightarrow e^{i q L} = -1$ and
$p\in\mathcal{S}_o\Leftrightarrow e^{2 i p (L+1)} = e^{2i\theta_p}$.
The sum over $j$ in \eqref{Eq:Mtilde} can be performed exactly and the summation over $p$ can be turned into a contour integral by means of the residue theorem \cite{longversion}.  
Correlations between fermions on the left side are related to those on the right side by chain inversion.
In addition, the correlations between fermions lying on different sides of the defect decay to zero.  We can thus focus on the right side.
If $\ell,n$ are such that $|\ell-n|\ll vt$, 
in the limit of large time we find~\cite{longversion}
\be
\Gamma_{\ell n}(t)\simeq \Gamma^{\textsc{lqss}}_{r/(vt),\ell-n}-\Gamma^{\textsc{b}}_{2 r}\,.
\label{Eq:solution}
\ee
where $r=(\ell+n)/2$,
 \begin{align}
&\Gamma^{\textsc{lqss}}_{\kappa,z}=\int_{-\pi}^\pi\!\frac{{\rm d}p}{2\pi}e^{-i z p}(f^\textsc{y}_p \tilde\sigma^y_{p}+f^\textsc{i}_{|p|} \tilde\sigma^y_{p}\theta_{|\varepsilon'_p|-\kappa v}+f^\textsc{i}_p \1 \theta_{\kappa v-|\varepsilon'_p|})\nonumber\\
&\Gamma^{\textsc{b}}_z= \int_{-\pi}^\pi\!\frac{{\rm d}p}{2\pi} e^{-i z p} (f^\textsc{i}_{|p|}+ f^\textsc{y}_p) \sigma^y e^{i\theta_p}\,
\label{Eq:Gb}
\end{align}
and $\theta_x$ is the step function that is nonzero and equal to $1$ only for $x>1$.
Expression \eqref{Eq:solution} exhibits the light-cone behavior discussed earlier, with $v=\text{max}_{p}|\varepsilon'_p|=J$ playing the role of the Lieb-Robinson velocity. For $n, \ell> v t(\gg1)$, it is as if there were no defect ($H^{(h)}_1\mapsto H^{(h)}_0$), while for $n,\ell<v t$ there is a nontrivial dependence on the position, which also persists in the limit of large $r$. 
To illustrate this, we show the energy density as a function of the distance from the boundary~--~Fig.~\ref{f:energy}.
From \eqref{Eq:solution} we can identify the correlation matrices $\Gamma^{\textsc{ness}}$ and $\Gamma^{\textsc{ness}}_+$ of $\rho^{\textsc{ness}}$ and $\rho_+^{\textsc{ness}}$ respectively 
\be\label{eq:Gness}
\Gamma^{\textsc{ness}}_{\ell n}=\Gamma^{\textsc{lqss}}_{0,\ell-n}+\Gamma_{\ell+n}^{\textsc{B}}\, ,\qquad \Gamma^{\textsc{ness}}_{+,\ell n}=\Gamma^{\textsc{lqss}}_{0,\ell-n}\, .
\ee
In particular, starting from the ground state of \eqref{eq:Hini}, for large enough $\Delta$ the initial state is no longer an eigenstate of chain inversion since $f^{\textsc{i}}_p\neq 0$. Therefore $ \Gamma^{\textsc{ness}}_{+,\ell n}$ \emph{differs} from the correlation matrix of the stationary state $\Gamma^{\textsc{GGE}}_{\ell n}=\Gamma^{\textsc{lqss}}_{1,\ell-n}$ describing correlations outside the light-cone. This is consistent with the picture presented: the boundary causes all the charges of the TFIC which are odd under reflection to become extinct~\cite{CLopenXY}. We explicitly verified that $\mathcal A_{\tilde Q}$~\eqref{eq:AQ} is non-zero when the GGE outside the light-cone is not reflection symmetric, \emph{i.e.} for $f^{\textsc{i}}_p\neq0$.

Fig.~\ref{f:energy} clearly shows that the energy in the NESS does not match the one in the GGE, despite $H_1$ being both in \eqref{eq:GGE} and in \eqref{eq:barrho}. This invalidates any attempt to construct the NESS by imposing the integrals of motion.

Importantly, in the space-time scaling limit with $\kappa=r/(vt)>0$,
the boundary part $\Gamma^{\rm B}_{2r}$ of \eqref{Eq:solution} approaches zero and the remaining part is the correlation matrix of a stationary state for the clean model. 
Since the equivalence holds only at $\kappa$ fixed and increasing the time changes the effective stationary state,  
we will refer to it as a \emph{locally-quasi-stationary state} and call it~$\rho_\kappa^{\textsc{lqss}}$.

\paragraph{Statistical description.}
\begin{figure}
\includegraphics[width=0.5\textwidth]{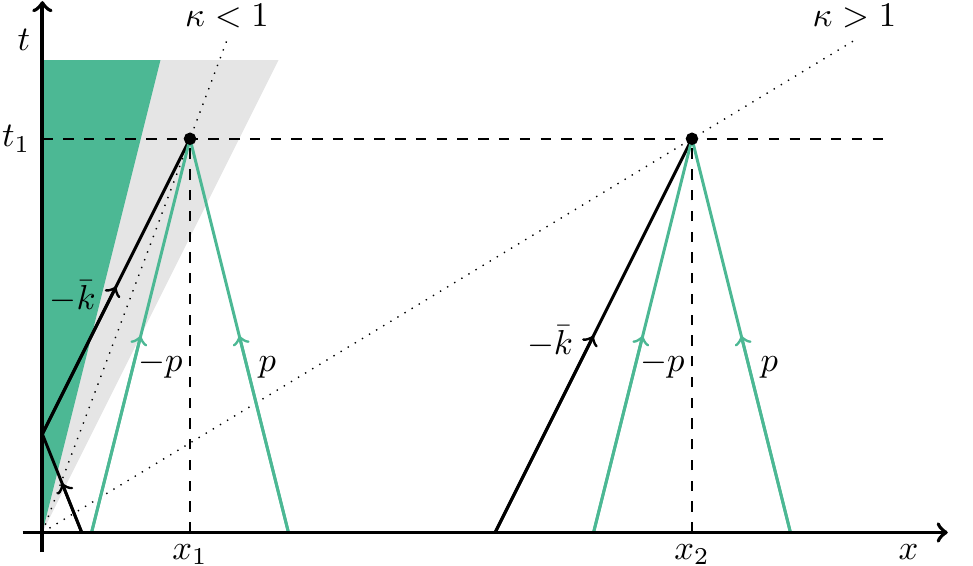}
\caption{Semiclassical interpretation of the invariants.
Arrows represent quasi-particle excitations and $\bar k$ corresponds to the maximal velocity. Circles indicate space-time points: one is inside the light-cone (gray triangle in background) and one outside. The green triangle on top is the light-cone for the quasi-particles with velocity smaller than $u_p<v$. The quasi-particle number is constant along the trajectory. }\label{f:lc}
\end{figure}
We now show that the general method proposed in the first part of this Letter can be used to obtain \eqref{Eq:solution} without solving the dynamics. 

First, we take $\rho^{\textsc{ness}}$ of the form \eqref{eq:NESS}. 
 Since the charges of the TFIC with open boundary conditions are quadratic forms of fermions~\cite{CLopenXY}, the NESS is Gaussian and completely characterized by its  correlation matrix $\Gamma^{\textsc{ness}}$.
As $\rho^{\textsc{ness}}$ commutes with $H_1$, 
$\sum_{\alpha,\beta}\sum_{\ell,n} [\Gamma^{\textsc{ness}}_{\ell n}]_{\alpha \beta} a_\ell^{\alpha} a_n^{\beta}$ is a linear combination of the charges. From the findings of \cite{CLopenXY} it then follows that
 \be
\Gamma^{\textsc{ness}}_{\ell n}=\int_{-\pi}^\pi\frac{{\rm d}p}{2\pi}g(p)(e^{i (n-\ell) p}\tilde\sigma^y_{p}-e^{-i (\ell+n) p} \sigma^y e^{i\theta_p}),
\label{Eq:OGGE}
\ee
where $g(p)=g(-p)$ is an arbitrary function. 

Our preliminary assumption \eqref{eq:NESS} is satisfied: \emph{the correlation matrix of the NESS in~\eqref{eq:Gness} is of the form~\eqref{Eq:OGGE}}. 

The next step is to construct invariants to fix $g(p)$. We can restrict our search to the densities $q_j{[f]}$ of the charges $Q[f]$ of the clean model such that $Q[f]=\sum\nolimits_p f(p)n_p\equiv \sum\nolimits_j q_j[f]$,
where $n_p$ are the occupation numbers commuting with $H_0$.
To be approximated by local operators, $q_j[f]$ must have tails decaying faster than the inverse of the width. The spreading properties of these observables can be deduced from the semiclassical picture of quasi-particle excitations, produced everywhere after the quench and moving freely~\cite{cc-05}.
In presence of a boundary this picture holds true as long as the quasi-particles are far enough from the boundary. When they hit it, they are scattered back and then move freely again~\cite{RI:sc11}. In this interpretation, $\mathrm{tr}[\rho_t q_j[f]]$  is a weighted measure of the number of quasi-particles reaching point $j$ at time $t$. 
The number of quasi-particles with a given momentum is constant along the trajectory  and, if the quasi-particle is moving towards the boundary, is the same as in the initial state. Being the initial state homogeneous, we conclude that the number of quasi-particles with velocity $u_p\geq 0$~\cite{f:3} is independent of both position and time and can be computed in the GGE outside the light-cone -- Fig.\,\ref{f:lc}. Thus, $g(p)$ in \eqref{Eq:OGGE} can be fixed by imposing the invariants $\mathrm{tr}[\rho^{\textsc{gge}}q_j[\sin(n p)\theta_{u_p}]]$, indeed in the TFIC $q_j[\sin(n p)\theta_{u_p}]$ can be approximated by local operators. This gives
$\mathrm{tr}[\rho^{\textsc{ness}}_+ n_p]=\mathrm{tr}[\rho^{\textsc{gge}}n_p]$ for $u_p\geq  0$
and leads to \eqref{eq:Gness}, proving the efficacy of our scheme. 

This is not the end of the story. 
We can also reconstruct the expectation values of local observables in the space-time scaling limit for any $\kappa=r/(v t)$. Quasiparticles with velocity $u_p>-\kappa v$ do not interact with the boundary and their number is the same as in the GGE outside the light-cone -- Fig.~\ref{f:lc}.
On the other hand, quasi-particles with velocity $u_p<-\kappa v$ hit the boundary in the past. 
We can compute their occupation numbers in $\rho^{\textsc{ness}}_+$ just after the collision. That is to say
\be\label{eq:theta}
\!\mathrm{tr}[\rho^{\textsc{lqss}}_\kappa n_p]=\theta_{u_p+\kappa v}\mathrm{tr}[\rho^{\textsc{gge}}n_p]+\theta_{-u_p-\kappa v}\mathrm{tr}[\rho^{\textsc{ness}}_+n_p].
\ee
Imposing  the GGE form  \eqref{eq:GGE}  to $\rho^{\textsc{lqss}}_\kappa$ produces \eqref{Eq:solution}. Remarkably, we have access to \emph{all} the late time correlations without the need of solving the dynamics! 
\paragraph{Comments and generalization. }

Equation \eqref{eq:theta} is the result of imposing  ``left'' and ``right'' invariants after a boost transformation at velocity $\kappa v$. Left (right) invariants  are weighted measures of the number of quasi-particles with negative (positive) velocities. 

This procedure is also effective in the interacting case, where the {LQSS} can be obtained by a slight modification of \eqref{eq:theta}.
The first step is to replace $\mathrm{tr}[\rho\, n_p]$ by the appropriate measures of the number of quasi-particles in the state $\rho$. When a thermodynamic Bethe ansatz description applies,  these are the so-called ``filling functions'' $\vartheta_{j}(\lambda)=\rho_{j}^{p}(\lambda)/(\rho^p_{j}(\lambda)+\rho_{j}^h(\lambda))$ \cite{Korepinbook}. Here $\rho_{j}^{p,h}(\lambda)$ describe the distributions of ``particles'' and ``holes'' of the $j$-th stable elementary excitation. Using the quasi-stationary properties of the {LQSS}, $\rho^{\textsc{lqss}}_\kappa$ is then characterized by $\kappa$-dependent filling functions $\vartheta_{\kappa, j}(\lambda)$. The second important point is that interactions cause the spreading velocities $u_{\kappa,p}$ to depend themselves on the (macro)state $\rho^{\textsc{lqss}}_\kappa$\cite{bonnes14}, so $\vartheta_{\kappa, j}(\lambda)$ must be computed in a self-consistent way.

The most problematic aspect is to determine the form of the NESS: \eqref{eq:theta} fixes its filling functions only for $u_{0,p}>0$. Doing this requires the knowledge of the charges with the defect. For a simple defect  switching off the interaction between two sites the filling functions are expected to be even \cite{openXYZ}. This allows to completely fix the {NESS}.
 
\paragraph{Conclusions.}
We have considered global quenches where the post-quench Hamiltonian has a localized defect. 
We proposed a procedure to determine the non-equilibrium steady state emerging from the impurity. This is based on the construction of invariants, \emph{i.e.} expectation values of observables that do not depend on the distance from the defect (in a particular limit, \emph{c.f.}~\eqref{eq:invq}). We applied the suggested scheme to a state evolving under the Hamiltonian of the TFIC with open boundary conditions; we checked our predictions against the analytic solution of the dynamics. 
A nontrivial NESS emerges only if the pre-quench Hamiltonian breaks reflection symmetry; however, this is specific to the TFIC and examples with symmetric pre- and post-quench Hamiltonians are known~\cite{Fdefect}.

Finally, we note that the same procedure can be used to reproduce the NESS and the LQSS also when two semi-infinite chains with different global properties are joined together~\cite{thermoelectricengine, BD:review, VM:review, twotemperatures,twomagnetizations,otherdifferentglobalproperties,conjTT}.

\begin{acknowledgments}
We thank Pasquale Calabrese, Fabian Essler, Andrea De Luca, Leonardo Mazza, Jacopo De Nardis, Lorenzo Piroli, and Neil Robinson for useful discussions. 
We thank the Isaac Newton Institute for Mathematical Sciences, under grant EP/K032208/1, for hospitality during the earliest stages of the collaboration.
This work was supported by the ERC under Starting Grant 279391 EDEQS (BB) and by LabEX ENS-ICFP:ANR-10-LABX-0010/ANR-10-IDEX-0001-02 PSL* (MF). 

\emph{Note added:}
The validity of the generalization to interacting models has been shown in Refs.~\cite{CAD:hydro,BCNF}, while this paper was under evaluation.

\end{acknowledgments}


\begin{thebibliography}{99}

\bibitem{vN} J. von Neumann, Z. Phys. \href{\doi10.1007/BF01339852}{\bf 57}, 30 (1929).


\bibitem{kww-06}
T. Kinoshita, T. Wenger,  and D. S. Weiss, 
 Nature \href{\doi10.1038/nature04693}{\bf 440}, 900 (2006).

\bibitem{exp} M. Greiner \emph{et al}, 
Nature \href{\doi10.1038/nature00968}{\bf 419}, 51-54 (2002); S. Hofferberth, I. Lesanovsky \emph{et al}, 
Nature \href{\doi10.1038/nature06149}{\bf 449}, 324-327 (2007); 
L. Hackermuller, U. Schneider \emph{et al},
Science \href{\doi10.1126/science.1184565}{\bf 327}, 1621 (2010);
S. Trotzky, Y.-A. Chen \emph{et al}, 
Nature Phys. \href{\doi10.1038/nphys2232}{\bf 8}, 325 (2012); M. Gring, M. Kuhnert \emph{et al}, 
Science \href{\doi10.1126/science.1224953}{\bf 337}, 1318 (2012):
U. Schneider, L. Hackerm\"uller \emph{et al}, 
Nature Phys. \href{\doi10.1038/nphys2205}{\bf 8}, 213 (2012); 
M. Cheneau, P. Barmettler \emph{et al}, 
Nature \href{http://dx.doi.org/10.1038/nature10748}{\bf 481}, 484 (2012); 
T. Langen, R. Geiger \emph{et al}, 
Nature Physics \href{http://dx.doi.org/10.1038/nphys2739}{\bf 9}, 640 (2013);
F. Meinert, M.J. Mark \emph{et al}, 
Phys. Rev. Lett. \href{http://dx.doi.org/10.1103/PhysRevLett.111.053003}{\bf 111}, 053003 (2013); T. Fukuhara, A. Kantian \emph{et al}, 
Nature Physics \href{\doi10.1038/nphys2561}{\bf 9}, 235 (2013); T. Fukuhara, P. Schau{\ss} \emph{et al}, 
Nature \href{\doi10.1038/nature12541}{\bf 502}, 76 (2013);
J.P. Ronzheimer, M. Schreiber \emph{et al}, 
Phys. Rev. Lett. \href{http://dx.doi.org/10.1103/PhysRevLett.110.205301}{\bf 110}, 205301 (2013);
P. Jurcevic, B. P. Lanyon \emph{et al}, 
Nature \href{http://dx.doi.org/10.1038/nature13461}{\bf 511}, 202 (2014).

\bibitem{EF:review} F.~H.~L.~Essler and M.~Fagotti, J. Stat. Mech. (2016)  \href{http://dx.doi.org/10.1088/1742-5468/2016/06/064002}{064002}.


\bibitem{GE15}
C.~Gogolin and J.~Eisert, Rep. Prog. Phys. \href{http://dx.doi.org/10.1088/0034-4885/79/5/056001}{\bf 79}, 056001 (2016).

\bibitem{P:review} E. Ilievski, M. Medenjak \emph{et al}, Stat. Mech. (2016) \href{http://dx.doi.org/10.1088/1742-5468/2016/06/064008}{064008}.

\bibitem{quench}
M. Rigol, A. Muramatsu, and M. Olshanii,  
Phys. Rev. A \href{http://dx.doi.org/10.1103/PhysRevA.74.053616} {\bf 74}, 053616 (2006);
M. A. Cazalilla, 
Phys. Rev. Lett. \href{http://dx.doi.org/10.1103/PhysRevLett.97.156403} {\bf 97}, 156403 (2006);
P. Calabrese and  J. Cardy,  
J. Stat. Mech. (2007) \href{http://dx.doi.org/10.1088/1742-5468/2007/06/P06008}{P06008};
M. Cramer, C. M. Dawson, J. Eisert, and T. J. Osborne, 
Phys. Rev. Lett. \href{http://dx.doi.org/10.1103/PhysRevLett.100.030602}{\bf 100}, 030602 (2008);
T. Barthel and U. Schollw\"ock, 
Phys. Rev. Lett. \href{http://dx.doi.org/10.1103/PhysRevLett.100.100601}{\bf 100}, 100601 (2008);
A. Silva, Phys. Rev. Lett. \href{http://dx.doi.org/10.1103/PhysRevLett.101.120603}{\bf 101}, 120603 (2008);
P. Calabrese, F. H. L. Essler and M. Fagotti,  
Phys. Rev. Lett. \href{http://dx.doi.org/10.1103/PhysRevLett.106.227203}{\bf 106}, 227203 (2011);
M. Fagotti and F.H.L. Essler, Phys.~Rev.~B \href{http://dx.doi.org/10.1103/PhysRevB.87.245107}{\bf 87}, 245107 (2013);
J.-S. Caux and R. M. Konik, 
Phys. Rev. Lett. \href{http://dx.doi.org/10.1103/PhysRevLett.109.175301}{\bf 109}, 175301 (2012);
F. H. L. Essler, S. Evangelisti, and M. Fagotti, 
Phys. Rev. Lett.  \href{http://dx.doi.org/10.1103/PhysRevLett.109.247206}{\bf 109}, 247206 (2012);
M. Collura, S. Sotiriadis, and P. Calabrese, 
Phys. Rev. Lett. \href{http://dx.doi.org/10.1103/PhysRevLett.110.245301}{\bf 110}, 245301 (2013);
J.-S.~Caux and F.H.L.~Essler, Phys. Rev. Lett. \href{http://dx.doi.org/10.1103/PhysRevLett.110.257203}{\bf 110}, 257203 (2013);
G. Mussardo, 
Phys. Rev. Lett. \href{http://dx.doi.org/10.1103/PhysRevLett.111.100401} {\bf 111}, 100401 (2013);
B. Pozsgay, 
J. Stat. Mech. (2013) \href{http://dx.doi.org/10.1088/1742-5468/2013/07/P07003}{P07003};
M. Fagotti and F. H. L. Essler, 
J. Stat. Mech. (2013) \href{http://dx.doi.org/10.1088/1742-5468/2013/07/P07012}{P07012};
M. Fagotti, M. Collura \emph{et al}, 
Phys. Rev. B \href{http://dx.doi.org/10.1103/PhysRevB.89.125101} {\bf 89}, 125101 (2014);
B. Wouters, J. De Nardis \emph{et al}, 
Phys. Rev. Lett.  \href{http://dx.doi.org/10.1103/PhysRevLett.113.117202}{\bf 113}, 117202 (2014);
B. Pozsgay, M. Mesty\'{a}n \emph{et al}, 
Phys. Rev. Lett. \href{http://dx.doi.org/10.1103/PhysRevLett.113.117203}{{\bf 113}}, 117203 (2014);
S. Sotiriadis and P. Calabrese, 
J. Stat. Mech. (2014) \href{http://dx.doi.org/10.1088/1742-5468/2014/07/P07024}{P07024};
G. Goldstein and N. Andrei, Phys. Rev. A \href{\doi10.1103/PhysRevA.90.043625}{\bf 90}, 043625 (2014);
F. H. L. Essler, G. Mussardo, and M. Panfil,  
Phys. Rev. A \href{http://dx.doi.org/10.1103/PhysRevA.91.051602}{\bf 91}, 051602(R); E. Ilievski, J. De Nardis \emph{et al}, 
Phys. Rev. Lett. \href{\doi10.1103/PhysRevLett.115.157201}{\bf 115}, 157201 (2015); E. Ilievski, E. Quinn  \emph{et al}, 
J. Stat. Mech. (2016) \href{http://dx.doi.org/10.1088/1742-5468/2016/06/063101}{063101}; L. Piroli, P. Calabrese, F.H.L. Essler, Phys. Rev. Lett. \href{\doi10.1103/PhysRevLett.116.070408}{\bf 116}, 070408 (2016); B. Bertini, L. Piroli, P. Calabrese, J. Stat. Mech. (2016) \href{http://dx.doi.org/10.1088/1742-5468/2016/06/063102}{063102}.



\bibitem{Rigol07}
M. Rigol, V. Dunjko, V. Yurovsky, and M. Olshanii,
Phys. Rev. Lett. \href{http://dx.doi.org/10.1103/PhysRevLett.98.050405}{\bf 98}, 050405  (2007).

\bibitem{Doyon} B. Doyon, 
arXiv:\href{http://arxiv.org/abs/1512.03713}{1512.03713} (2015).

\bibitem{thermalization}
J. M. Deutsch, 
Phys. Rev. A
\href{http://dx.doi.org/10.1103/PhysRevA.43.2046}{\bf 43}, 2046
(1991);   M. Srednicki, 
Phys. Rev. E
  \href{http://dx.doi.org/10.1103/PhysRevE.50.888}{\bf 50}, 888
  (1994);
  M. Rigol, V. Dunjko, and 
M. Olshanii, 
Nature \href{\doi10.1038/nature06838}{\bf 452}, 854 (2008);
  M. Rigol and M. Srednicki, 
 Phys. Rev. Lett. \href{http://dx.doi.org/10.1103/PhysRevLett.100.100601}{\bf 108}, 110601 (2012).


\bibitem{prethermalization}
C. Kollath, A. M. L\"auchli, and E. Altman, 
Phys. Rev. Lett. \href{http://dx.doi.org/10.1103/PhysRevLett.98.180601}{\bf 98}, 180601 (2007);      
M. Moeckel and S. Kehrein, 
Phys. Rev. Lett. \href{http://dx.doi.org/10.1103/PhysRevLett.100.175702}{\bf 100}, 175702 (2008);
Ann. of Phys. \href{\doi10.1016/j.aop.2009.03.009}{\bf 324}, 2146 (2009);
 M. Kollar, F.A. Wolf, and M. Eckstein, 
Phys. Rev. B \href{http://dx.doi.org/10.1103/PhysRevB.84.054304}{\bf 84}, 054304 (2011);
M. Stark and M. Kollar, 
arXiv:\href{http://arxiv.org/abs/1308.1610}{1308.1610} (2013);
M. Marcuzzi, J. Marino \emph{et al}, 
Phys. Rev. Lett. \href{http://dx.doi.org/10.1103/PhysRevLett.111.197203}{\bf 111}, 197203 (2013);
A. Mitra, 
Phys. Rev. B \href{\doi10.1103/PhysRevB.87.205109}{\bf 87}, 205109 (2013);
F.H.L. Essler, S. Kehrein \emph{et al}, 
Phys. Rev. B \href{http://dx.doi.org/10.1103/PhysRevB.89.165104}{\bf 89}, 165104 (2014);
 A. Chiocchetta, M. Tavora \emph{et al}, 
Phys. Rev. B \href{http://dx.doi.org/10.1103/PhysRevB.91.220302}{\bf 91}, 220302(R) (2015); Erratum Phys. Rev. B \href{http://dx.doi.org/10.1103/PhysRevB.92.219901}{\bf 92}, 219901 (2015); 
 B. Bertini, F.H.L. Essler \emph{et al}, 
Phys. Rev. Lett. \href{http://dx.doi.org/10.1103/PhysRevLett.115.180601}{\bf 115}, 180601 (2015); 
G.P. Brandino, J.-S. Caux, and R.M. Konik, 
Phys. Rev. X \href{http://dx.doi.org/10.1103/PhysRevX.5.041043}{\bf 5}, 041043 (2015);
 M. Babadi, E. Demler, and M. Knap, 
Phys. Rev. X \href{http://dx.doi.org/10.1103/PhysRevX.5.041005}{\bf 5}, 041005 (2015);
 N. Nessi and A. Iucci, 
arXiv:\href{http://arxiv.org/abs/1503.02507}{1503.02507} (2015).

\bibitem{prerelaxation}
M. Fagotti, 
J. Stat. Mech. (2014) \href{ttp://dx.doi.org/10.1088/1742-5468/2014/03/P03016}{P03016};
 B. Bertini and M. Fagotti, 
J. Stat. Mech. (2015) \href{http://dx.doi.org/10.1088/1742-5468/2015/07/P07012}{P07012};
 M. Fagotti and M. Collura, 
arXiv:\href{http://arxiv.org/abs/1507.02678}{1507.02678} (2015).

\bibitem{LGS:review} T.~Langen, T.~Gasenzer, and J.~Schmiedmayer, 
 J. Stat. Mech. (2016) \href{1http://dx.doi.org/10.1088/1742-5468/2016/06/064009}{064009}.

\bibitem{prethermalizationexp}
 M. Gring, M. Kuhnert \emph{et al}, 
Science \href{\doi10.1126/science.1224953}{\bf 337}, 1318 (2012);
T. Langen, S. Erne \emph{et al}, 
Science \href{\doi10.1126/science.1257026}{\bf  348}, 207 (2015).

\bibitem{thermoelectricengine} J.-P. Brantut, C. Grenier \emph{et al}, 
Science \href{\doi10.1126/science.1242308}{\bf 342}, 713 (2013).

\bibitem{BD:review} D. Bernard, B. Doyon, J. Stat. Mech. (2016) \href{\doi10.1088/1742-5468/2016/06/064005}{064005}.

\bibitem{VM:review} R. Vasseur, J. E. Moore, J. Stat. Mech. (2016) \href{\doi10.1088/1742-5468/2016/06/064010}{064010}.


\bibitem{twotemperatures}
H. Spohn, J. L. Lebowitz, 
Comm. Math. Phys., \href{\doi10.1007/BF01614132}{\bf 54}, 97 (1977);    
T. Platini and D. Karevski, 
  J. Phys. A: Math. Theor. \href{http://dx.doi.org/10.1088/1751-8113/40/8/002}{\bf 40} 1711 (2007);
W.H. Aschbacher and C.-A. Pillet, 
J. Stat. Phys. \href{\doi10.1023/A:1024619726273}{\bf 112}, 1153 (2003);
W.H. Aschbacher and J.-M. Barbaroux, 
 Lett. Math. Phys. \href{\doi10.1007/s11005-006-0049-7}{\bf 77}, 11 (2006);
  D. Bernard and B. Doyon, 
  J. Phys. A: Math. Theor. \href{http://dx.doi.org/10.1088/1751-8113/45/36/362001}{\bf 45}, 362001 (2012);
 M. Mintchev, J. Phys. A: Math. Theor. \href{\doi10.1088/1751-8113/44/41/415201}{\bf 44} 415201 (2011).
A. De Luca, J. Viti \emph{et al}, 
 Phys. Rev. B \href{http://dx.doi.org/10.1103/PhysRevB.88.134301}{\bf 88}, 134301 (2013);
C. Karrasch, R. Ilan, and J. E. Moore, 
Phys. Rev. B \href{http://dx.doi.org/10.1103/PhysRevB.88.195129}{\bf 88}, 195129 (2013);
M. Mintchev and P. Sorba, 
 J. Phys. A: Math. Theor. \href{http://dx.doi.org/10.1088/1751-8113/46/9/095006}{\bf 46}, 095006 (2013);
B. Doyon, M. Hoogeveen, and D. Bernard, 
 J. Stat. Mech. (2014) \href{http://dx.doi.org/10.1088/1742-5468/2014/03/P03002}{P03002}; 
A. De Luca, G. Martelloni, and J. Viti, 
 Phys. Rev. A \href{http://dx.doi.org/10.1103/PhysRevA.91.021603}{\bf 91}, 021603(R) (2015);
B. Doyon, A. Lucas \emph{et al}, 
  J. Phys. A: Math. Theor. \href{http://dx.doi.org/10.1088/1751-8113/48/9/095002}{48} 095002 (2015);
B. Doyon, 
  Nucl. Phys. B \href{\doi10.1016/j.nuclphysb.2015.01.007}{\bf 892}, 190 (2015);
A. Biella, A. De Luca \emph{et al}, 
Phys. Rev. B \href{\doi10.1103/PhysRevB.93.205121}{\bf 93}, 205121 (2016).

\bibitem{conjTT}O. Castro-Alvaredo, Y. Chen \emph{et al}, 
  J. Stat. Mech. (2014) \href{http://dx.doi.org/10.1088/1742-5468/2014/03/P03011}{P03011};
  A. De Luca, J. Viti \emph{et al}, 
Phys. Rev. B \href{http://dx.doi.org/10.1103/PhysRevB.90.161101}{\bf 90}, 161101(R) (2014).

\bibitem{twomagnetizations} 
T. Antal, Z. Racz \emph{et al}, 
Phys. Rev. E \href{http://dx.doi.org/10.1103/PhysRevE.59.4912}{\bf 59}, 4912 (1999);
T. Sabetta and G. Misguich, 
  Phys. Rev. B \href{http://dx.doi.org/10.1103/PhysRevB.88.245114}{\bf 88}, 245114 (2013).


\bibitem{otherdifferentglobalproperties}
S. Sotiriadis and J. Cardy, 
J. Stat. Mech. \href{\doi10.1088/1742-5468/2008/11/P11003}{P11003} (2008);
P. Calabrese, C. Hagendorf, and P. Le Doussal, 
J. Stat. Mech. \href{\doi10.1088/1742-5468/2008/07/P07013}{P07013} (2008);
  J. Lancaster and A. Mitra, 
  Phys. Rev. E \href{http://dx.doi.org/10.1103/PhysRevE.81.061134}{\bf 81}, 061134 (2010);
V. Eisler and Z. Racz, 
Phys. Rev. Lett. \href{http://dx.doi.org/10.1103/PhysRevLett.110.060602}{\bf
  110}, 060602 (2013); 
J. Viti, J.-M. St\'ephan \emph{et al}, 
  arXiv:\href{http://arxiv.org/abs/1507.08132}{1507.08132} (2015). 
  

  
\bibitem{Ruelle} D. Ruelle, J. Stat. Phys. \href{\doi10.1023/A:1018618704438}{\bf 98}, 57 (2000).
  

  \bibitem{BD:nessCFT} D. Bernard and B. Doyon, 
  Ann. H. Poincar\'e \href{\doi10.1007/s00023-014-0314-8}{\bf 16}, 113 (2015).

\bibitem{Fdefect} 
M. Fagotti, 
arXiv:\href{http://arxiv.org/abs/1508.04401}{1508.04401} (2015).

\bibitem{SM:toc} M. Schir\`o and A. Mitra, 
Phys. Rev. Lett. \href{http://dx.doi.org/10.1103/PhysRevLett.112.246401}{\bf 112}, 246401 (2014).

\bibitem{CVZ:14}
C.-C. Chien, M. Di Ventra, M. Zwolak, 
Phys. Rev. A \href{\doi10.1103/PhysRevA.90.023624}{90}, 023624 (2014).

\bibitem{f:1} This readily follows from the identify  $\mathcal O_A(t)=e^{i H_0 t}\mathcal O_Ae^{-i H_0 t}+i\int_0^t\mathrm d \tau e^{i H_0\tau}[\hat d,\mathcal O_A(t-\tau)]e^{-i H_0\tau}$.

\bibitem{openXYZ} M.P. Grabowski and P. Mathieu, J. Phys. A: Math. Gen. \href{http://dx.doi.org/10.1088/0305-4470/29/23/024}{\bf 29} 7635 (1996).

\bibitem{LR72}
E. H. Lieb and D. W. Robinson, 
Commun. Math. Phys. \href{http://dx.doi.org/10.1007/BF01645779}{\bf 28}, 251
(1972).

\bibitem{bravyi06}
S. Bravyi, M. B. Hastings, and F. Verstraete, 
Phys. Rev. Lett.
\href{http://dx.doi.org/10.1103/PhysRevLett.97.050401}{\bf 97}, 050401 (2006).

\bibitem{f:2} 
This fixes the NESS only if there are no charges with vanishing bulk part. Otherwise, one should also impose the conservation of those special charges. 

\bibitem{cc-05}   
P. Calabrese and  J. Cardy, 
J. Stat. Mech. (2005) \href{http://dx.doi.org/10.1088/1742-5468/2005/04/P04010}{P04010}. 




\bibitem{longversion}
B. Bertini and M. Fagotti, \emph{in preparation}.


\bibitem{CLopenXY} M. Fagotti, 
J. Stat. Mech. (2016) \href{\doi10.1088/1742-5468/2016/06/063105}{063105}.


\bibitem{RI:sc11} H. Rieger and F. Igl\'oi, 
Phys. Rev. B \href{http://dx.doi.org/10.1103/PhysRevB.84.165117}{\bf 84}, 165117 (2011).



\bibitem{f:3}
By convention, operators that spread out to the right (in the Heisenberg picture) have positive velocity and, therefore, the quasi-particles produced after the quench with positive velocity move to the left.  


\bibitem{Korepinbook}
V.E. Korepin, A.G. Izergin, and N.M. Bogoliubov, {\em {Quantum Inverse
  Scattering Method, Correlation Functions and Algebraic Bethe Ansatz}}
  (Cambridge University Press, 1993).


\bibitem{bonnes14}
L. Bonnes, F.H.L. Essler and A. M. L\"auchli, 
Phys. Rev. Lett. \href{http://dx.doi.org/10.1103/PhysRevLett.113.187203}{\bf 113}, 187203 (2014).


\bibitem{CAD:hydro}
O.~A.~Castro-Alvaredo, B.~Doyon, and T.~Yoshimura, arXiv:\href{http://arxiv.org/abs/1605.07331}{1605.07331} (2016). 

\bibitem{BCNF}
B. Bertini, M. Collura \emph{et al}, 
 arXiv:\href{http://arxiv.org/abs/1605.09790}{1605.09790} (2016). 

\end{thebibliography}
\end{document}